\documentstyle[preprint,aps,psfig]{revtex}
\begin{document}

\draft

\title{Resonances in the current-voltage characteristics of a
dissipative Josephson junction}
\author{J. M\"ullers and A. Schmid}
\address{Institut f\"ur Theorie der Kondensierten Materie,
Universit\"at Karlsruhe, 76128 Karlsruhe, Germany}
\date{\today}

\maketitle

\begin{abstract}
The I-V characteristics of a Josephson junction shunted by an Ohmic resistor
shows sharp peaks when levels in neighbouring wells are crossing.
We consider the shape and size of these peaks using a double-well model
where the wells are given in parabolic approximation. The friction of
arbitrary strength is included by the help of a bath of infinitely many
degrees of freedom.
\end{abstract}



\section{Introduction}

The past decade has seen a considerable interest and remarkable
activity in an area which presently is often referred to as  macroscopic
quantum mechanics.  Specifically, one has been interested in quantum
phenomena of macroscopic objects \cite{Leggett86}.

In particular, macroscopic quantum tunneling \cite{Caldeira}
(quantum decay of a metastable state), and quantum coherence
\cite{Leggett87} have
been studied.  Soon, it became clear that dissipation has a profound
influence on these quantum phenomena.  Phenomenologically, dissipation
is the consequence of an interaction of the object with an environment
which can be thought of as consisting of infinitely many degrees
of freedom.  Specifically, the environmental degrees of freedom
may be chosen to be harmonic oscillators such that we may consider
the dissipation as a process where excitations, that are phonons, are
emitted and absorbed.  Such a model (Caldeira-Leggett model) has
been the basis in Ref. \cite{Caldeira} where the influence of dissipation on
tunneling has been explored.

As far as quantum coherence is concerned, the most simple system is an
object with two different quantum states: it is thought to
represent the limiting case of an object in a double-well potential
where only the lowest energy states in each of the two wells is relevant and
where the tunneling through the separating barrier allows for
transitions that probe the coherence.  Since a two-state system is
equivalent to a spin-one-half problem, this standard system is
often referred to by this name.  In particular, with the standard
coupling to a dissipative environment made of harmonic
oscillators, it is called the spin-boson problem which has been
studied repeatedly in the past \cite{Leggett87,Weiss}.

Level quantization and resonant tunneling have been observed recently
\cite{vaart} in a double-well quantum-dot system. However, the influence of
dissipation was not considered in this experiment.
On the other hand, it seems that Josephson junctions
are also suitable systems for obtaining experimental evidence pertaining
to macroscopic quantum effects.  In this context, evidence for level
quantization \cite{Martinis}, \cite{Larkin} and for quantum decay
\cite{Devoret} have been obtained; however, macroscopic quantum
coherence has evaded observation so far.

Specifically, a Josephson junction may be
 characterized by a current-phase relation
\begin{equation}
    I(\varphi) = I_J\sin{\varphi}\,,
\label{I.1}
\end{equation}
where the phase $\varphi$ is related to the voltage difference
$U$ by
\begin{equation}
   \hbar\dot{\varphi} = 2eU\,.
\label{I.2}
\end{equation}
Therefore, the phase of a Josephson junction shunted by a capacitance
$C$ and biased by an external current $I_x$ obeys a classical type of
equation of motion
\begin{equation}
   M\ddot{\varphi} = - \frac{\partial V(\varphi)}{\partial\varphi}\,,
\label{I.3}
\end{equation}
with the mass
\begin{equation}
   M = \left(\frac{\hbar}{2e}\right)^2 C
\label{I.4}
\end{equation}
and the potential energy
\begin{equation}
   V(\varphi) = - \frac{\hbar}{2e}\,
   \left[I_J \cos{\varphi} + I_x\varphi\right]\,.
\label{I.5}
\end{equation}

A widely discussed model of a dissipative object is the one where the
Josephson junction is also shunted by an Ohmic
resistor $R$.  In this case, the classical equation of motion
(\ref{I.3}) has to be replaced by
\begin{eqnarray}
   M\ddot{\varphi} &=& - \frac{\partial V(\varphi)}{\partial\varphi}
   -\eta\dot{\varphi} \,,\nonumber \\
   \eta &=& \left(\frac{\hbar}{2e}\right)^2 \frac{1}{R}\,.
\label{I.6}
\end{eqnarray}
The model of a dissipative environment according to the above
specification has been discussed by Caldeira and Leggett \cite{Caldeira}.

The potential energy $V(\varphi)$ of Eq. (\ref{I.5}) displays wells
at $\varphi \simeq 2 n\pi$ with depth shifted by an amount
$\Delta\simeq (2\pi\hbar/2e)I_x$.  If the wells are sufficiently
deep, one needs to concentrate only on transitions between pairs of
adjacent wells.  Thus, one arrives at the double well problem
mentioned above.

The analysis in this paper goes beyond the limiting situation
where only the lowest level in each of the two wells is of importance.
Roughly, this is realized when the level separation $\hbar(2E_J/M)^{1/2}\simeq
(2e\hbar I_J/C)^{1/2}$ is smaller than or comparable with $\Delta$.  In
particular, we will concentrate on resonance phenomena which are
expected to show up whenever two levels in the adjacent wells
happen to cross when the bias current $I_x$, that is $\Delta$,
is varied.

For such values of the bias current,
there appear sharp asymmetric peaks in the
current-voltage characteristic of the Josephson junction.  This phenomenon
has been studied by one of us (A.S., together with A.I. Larkin and
Yu. N. Ovchinnikov) \cite{Larkin88} within
the standard model in the one-phonon approximation.  For bias currents
that correspond to  crossings of the next and next nearest levels (e.g.
ground state in the left well and the first or second excited state at the
right side), it is possible to neglect processes in the reverse direction
provided that the temperature is sufficiently low.  Thus, the
restriction to a double well system receives additional support.

The transfer of the object from the left to the right potential well
is accompanied by the emission of an infinite number of phonons.
Therefore, in a recent paper by Ovchinnikov and one of us
(A.S.) \cite{Schmid94} the fact is taken into account that
in the resonance region, the contribution of phonons of small energy
is important as well as the contribution of resonance phonons
with energy equal to the distance between levels in the wells.

The above analysis relies on a perturbation theory for weak
dissipation. Moreover, the calculations in that article are rather
intricate. Therefore, we want to reconsider the problem in a more
transparent way and
thereby get rid of the restriction of small dissipation.

\section{Description of the model Hamiltonian and its eigenstates}
\label{sec:des}

\begin{figure}
\centerline{\psfig{figure=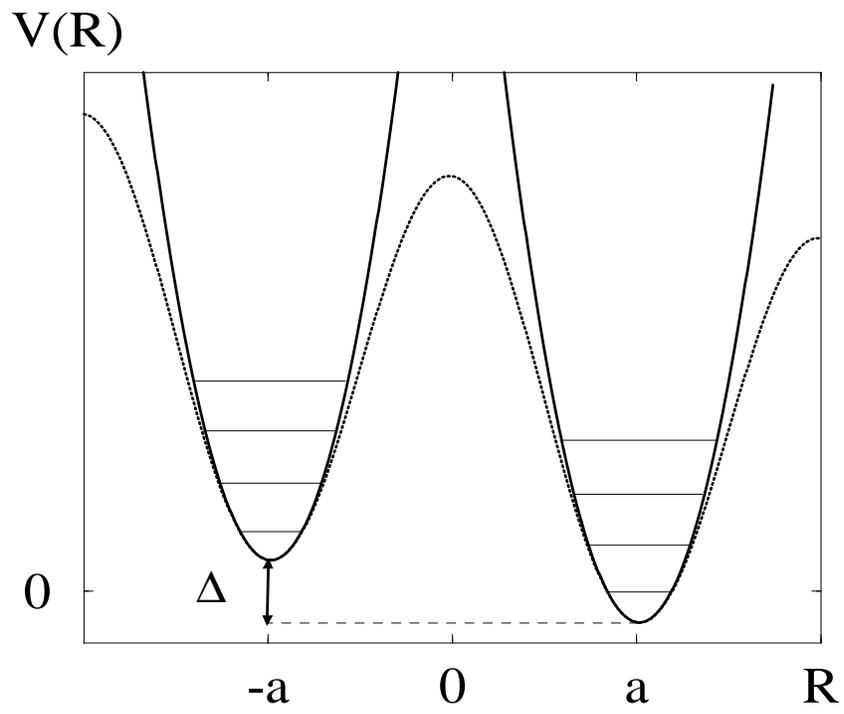,height=10cm,width=13cm}}
\caption{Double well potential with shift $a$ and bias $\Delta$}
\label{general}
\end{figure}

The model consists of a particle, called "'object"' (coordinate
$R_1$), which is coupled (in the sense of Caldeira and Leggett \cite{Caldeira})
to a "'bath"' of harmonic oscillators (coordinates $R_j$). We  shall use
the conventions $j \in \{2,\dots,N\}$ for the bath oscillators and
$k \in \{1,\dots,N\}$ for the indices of all coordinates in the model.
The double-well potential is approximated by two parabolas about the
minima of the two wells as it is depicted in fig. \ref{general}.

The phase $\varphi$ of the Josephson contact then corresponds to the
object coordinate $R_1$ of the model, and the voltage $U$ is related to
the tunneling rate $J$ by $2e U = \dot{\varphi} = 2\pi J$. As it has
already been remarked, the current $I_x$ is proportional to the bias $\Delta$
of the two wells. Thus, calculating the transition rate for different
values of the bias $\Delta$ is equivalent to the determination of
the I-V characteristics.

Specifically, we want to write the Hamiltonian of the model in the
form
\begin{eqnarray}
	\hat{H} & = & \frac{1}{2 m} \sum_k \hat{p}_k^2 +
	\hat{v}(\hat{R}_1) + \frac{m}{2} \sum_j \omega_j^2
	(\hat{R}_j-\hat{R}_1)^2\,,\nonumber\\
	\hat{v}(\hat{R}_1) & \approx & \frac{m}{2} \sum_\pm
	\Omega^2 (\hat{R}_1 \pm a)^2 \pm \frac{\Delta}{2}\,.
\label{II.1}
\end{eqnarray}
The states for the two situations "'object in the left well"'
and "'object in the right well"' will be denoted by $|\Lambda_L, L\rangle$
and $|\Lambda_R, R\rangle$, respectively. If one projects onto the
eigenstates $|n\rangle$ of the one-dimensional harmonic oscillator
and takes into account the shift of the wells, one arrives at the
following decomposition ($\varphi_n (R) = \langle R|n\rangle$):
\begin{eqnarray}
\label{II.2}
	\langle n_L, \{R_j\} | \Lambda_L, L\rangle & =: & \int dR_1\,
	\varphi_{n_L} (R_1+a)\, \phi^L_{\Lambda_L} (\{R_k\})\,,\\
\label{II.3}
	\langle n_R, \{R_j\} | \Lambda_R, R\rangle & =: & \int dR_1\,
	\varphi_{n_R} (R_1-a)\, \phi^R_{\Lambda_R} (\{R_k\})\,.
\end{eqnarray}

The situations "'object on the left"' and "'object on the right"'
differ only by the shift and the bias of the wells. Therefore,
one can find a unified representation by noting that
$\phi^L_{\Lambda} (\{R_k\}) = \Phi_{\Lambda} (\{R_k+a\})$ and
$\phi^R_{\Lambda} (\{R_k\}) = \Phi_{\Lambda} (\{R_k-a\})$.
The eigenstates $\Phi_\Lambda$ are defined by the relations
\begin{eqnarray}
	\Phi_\Lambda (\{R_k\}) & := & \langle \{R_k\} | \Lambda\rangle\,,\\
	\hat{H}_0 |\Lambda\rangle & = & E_\Lambda |\Lambda\rangle\,,\\
	\hat{H}_0 & = & \frac{1}{2 m} \sum_k \hat{p}_k^2 + \frac{m}{2}
	\sum_j \omega_j^2 (\hat{R}_j-\hat{R}_1)^2 +
	\frac{m}{2}\, \Omega^2 \hat{R}_1^2\,.
\end{eqnarray}
Thus, it follows from eqs. (\ref{II.2}) and (\ref{II.3}), respectively, that
\begin{eqnarray}
\label{II.4}
   \langle n_L, \{R_j\} | \Lambda_L, L\rangle &=&
   \langle n_L, \{R_j\} | \exp{(i a \sum_j \hat{p}_j)}\, |\Lambda_L>\,,\\
\label{II.5}
   \langle n_R, \{R_j\} | \Lambda_R, R\rangle &=&
   \langle n_R, \{R_j\} | \exp{(-i a \sum_j \hat{p}_j)}\, |\Lambda_R>\,,
\end{eqnarray}
where we have used the shift property of the momentum operator $\hat{p}$.

The coupling of the two wells is taken into account by means of
a tunneling Hamiltonian $\hat{H}_T$ which we represent in the form
\begin{eqnarray}
	\langle\Lambda_L, L | \hat{H}_T | \Lambda_R, R\rangle & = &
	\int d\{R_j\}\, \sum_{n_Ln_R} T_{n_Ln_R}\,
        \langle \Lambda_L, L | n_L, \{R_j\} \rangle \langle
        n_R, \{R_j\} | \Lambda_R, R\rangle\,.
\end{eqnarray}
Using again the momentum operator, one can write
\begin{equation}
   |x\rangle\langle x'| = e^{i\hat{p} (x'-x)}\, |x'\rangle\langle x'|
   = e^{i\hat{p} (x'-x)}\, \delta(x'-\hat{x})\,.
\end{equation}
 From this, we conclude that
\begin{eqnarray}
\label{II.6}
	\langle\Lambda_L, L | \hat{H}_T | \Lambda_R, R\rangle & = &
        \sum_{n_Ln_R} T_{n_Ln_R}\, \int dR_1 dR'_1 \,\frac{dQ}{2\pi}\,
        \varphi_{n_R}^{\ast} (R'_1) \,\varphi_{n_L} (R_1)\,\nonumber\\
        & & \times\, \langle \Lambda_L, L |
        e^{i\hat{p}_1 (R'_1-R_1)} e^{iQ (R'_1-\hat{R}_1)}
        | \Lambda_R, R\rangle\,.
\end{eqnarray}

\section{Transition Rate}
\label{sec:trans}

The net transition rate from the left well to the right one
is then in second order perturbation theory given by
\begin{eqnarray}
	J & = & 2\pi Z_0^{-1} \sum_{{\Lambda_L},{\Lambda_R}}
	|\langle\Lambda_L, L | \hat{H}_T | \Lambda_R, R\rangle|^2\,
	\,\delta (E_{\Lambda_L} - E_{\Lambda_R} + \Delta)\nonumber\\
	& & \times\, [e^{-\beta E_{\Lambda_L}} - e^{\beta E_{\Lambda_R}}]\,,
\end{eqnarray}
where $Z_0 = \mbox{Tr} \exp{(-\beta H_0)}$.
The $\delta$-function may be written in Fourier representation, and the
fact that the $E_\Lambda$ are eigenenergies of $\hat{H}_0$ serves us to
incorporate the energy conservation into Heisenberg time-dependent
operators $\hat{A}(t) = \exp{(i\hat{H}_0 t)} \hat{A} \exp{(-i\hat{H}_0 t)}$,
i.e.
\begin{eqnarray}
\label{III.0}
        \lefteqn{\langle\Lambda_L, L | \hat{H}_T | \Lambda_R, R\rangle\;
        \delta (E_{\Lambda_L} - E_{\Lambda_R} + \Delta) = }\nonumber\\
        & & = \int dt\, e^{i \Delta t}\,
        \langle\Lambda_L | e^{-i a \sum_k \hat{p}_k (t)} \,\hat{H}_T (t) \,
        e^{-i a \sum_k \hat{p}_k (t)} | \Lambda_R\rangle\,.
\end{eqnarray}

Then, collecting our results from eqs. (\ref{II.4}), (\ref{II.5}),
(\ref{II.6}), and (\ref{III.0}), we arrive at the expression
\begin{eqnarray}
	J & = & Z_0^{-1} (1-e^{-\beta\Delta})\, \int dt\, e^{i\Delta t}\,
	\sum_{n_L, n_R} \sum_{\overline{n}_L, \overline{n}_R}
	T_{n_L, n_R} T_{\overline{n}_L, \overline{n}_R}^\ast\,\nonumber\\
	& & \times\,\int \frac{dQ d\overline{Q}}{(2\pi)^2}\,
        \int dR_1 dR'_1 d\overline{R}_1
	d\overline{R}'_1\, \varphi_{n_L} (R_1) \,\varphi_{n_R}^\ast (R'_1)
	\,\varphi_{\overline{n}_L}^\ast (\overline{R}'_1)
	\,\varphi_{\overline{n}_R} (\overline{R}_1)\,\nonumber\\
	& & \times\, \mbox{Tr}\, \{e^{-\beta \hat{H}_0}\,
	e^{-2ia \sum_k \hat{p}_k(t)} \,e^{i\hat{p}_1(t) (R'_1 - R_1 + 2a)}
	\,e^{-iQ (\hat{R}_1(t) - R'_1)} \,\nonumber\\
	& & \times\,
	e^{+2ia \sum_k \hat{p}_k} \,e^{i\hat{p}_1 (\overline{R}'_1 -
        \overline{R}_1 - 2a)}\,
	e^{-i\overline{Q} (\hat{R}_1 - \overline{R}'_1)} \}\,.
\end{eqnarray}
Let us now use the relation
\begin{equation}
    	e^{-i (\hat{H}_0 + \hat{W}) t} =
	e^{-i \hat{H}_0 t} \,\hat{T}\,
	 e^{-i \int_0^t dt'\, \hat{W} (t')}
\end{equation}
which holds for $t>0$ when $\hat{T}$ is the time-ordering operator
and for $t<0$ when the anti time-ordering is used. If we define
$\langle \hat{A} \rangle := \mbox{Tr}\,\exp{(-\beta \hat{H}_0)} \hat{A}
/ Z_0$, we can write the following result for the transition rate:
\begin{eqnarray}
	J & = & (1-e^{-\beta\Delta})\, \int dt\,
        e^{i(\Delta - 2m\Omega^2 a^2)t}\,
	\sum_{n_L, n_R} \sum_{\overline{n}_L, \overline{n}_R}
	T_{n_L, n_R} T_{\overline{n}_L, \overline{n}_R}^\ast\,\nonumber\\
	& \times & \int \frac{dQ d\overline{Q}}{(2\pi)^2}\, \int dR dR'
        d\overline{R} d\overline{R}'\,
        e^{i Q \frac{R+R'}{2} + i \overline{Q} \frac{\overline{R}+
        \overline{R}'}{2}}\nonumber\\
        & \times & \varphi_{n_L} (R) \,\varphi_{n_R}^\ast (R'-2a)
	\,\varphi_{\overline{n}_L}^\ast (\overline{R}')
	\,\varphi_{\overline{n}_R} (\overline{R}-2a)\,\nonumber\\
	& \times & \bigg\langle \hat{T}\,
		\exp{[-i Q \hat{R}_1 (t) + i \hat{p}_1 (t) (R'-R)
		+ 2i m\Omega^2 a \int_0^t dt'\, \hat{R}_1 (t')}
		\nonumber\\
	& &  - i \overline{Q} \hat{R}_1 (0) + i \hat{p}_1 (0)
                 (\overline{R}' - \overline{R})]
	\bigg\rangle\,.
\label{III.res}
\end{eqnarray}

We are now in the position to make use of the fact that the Hamiltonian
is quadratic in all coordinates so that we can evaluate exactly
\begin{eqnarray}
\label{III.quad}
	\langle \hat{T} e^{i \int dt'\, \eta (t')
        \hat{R}_1 (t')} \rangle & = &
	e^{-\frac{i}{2} \int\int dt' dt'' \eta(t') D (t',t'') \eta(t'')}\,,\\
	D (t',t'') & := & -i \langle \hat{T}
        \hat{R}_1 (t') \hat{R}_1 (t'') \rangle\,.
\end{eqnarray}
By comparison with the last two lines in eq. (\ref{III.res}), the function
$\eta(t')$ is given by
\begin{eqnarray}
   \eta(t') &=& -Q \delta(t'-t) - \overline{Q} \delta(t') + 2m \Omega^2 a
   [\Theta(t')-\Theta(t'-t)]\nonumber\\
   & & + m(R-R') \delta'(t'-t)
   + m(\overline{R}-\overline{R}') \delta'(t')\,.
\end{eqnarray}
$\Theta(t)$ is meant to represent the step function.
The derivatives of the $\delta$-function arise from a partial integration
of terms containing $\hat{p}(t) = m d\hat{x}(t)/dt$. Note, that these
act only on the coordinates but not on the step functions which arise
due to the time ordering.

Moreover, the degrees of freedom of the
bath can be integrated out in the usual way
\cite{Caldeira} leading to a dissipative influence on the object.
One is then lead to the following form of the Fourier transform
of $D(t,t') \equiv D(t-t')$:
\begin{eqnarray}
\label{III.D}
        D(\omega) & = & \frac{D^R(\omega)}{1-\exp{(-\hbar\omega/k_BT)}}
        + \frac{D^R(-\omega)}{1-\exp{(\hbar\omega/k_BT)}}\,,\\
	{(D^R)}^{-1}(\omega) & = & m [(\omega+i0)^2 - \Omega^2] + i\eta\omega\,,
\end{eqnarray}
where we will use a spectral density $J(\omega) =
\eta \omega$, $0\leq\omega\leq\omega_c$ for the bath oscillators
(see also \cite{RMS} and appendix \ref{app:green}).

 From eqs. (\ref{III.res}) and (\ref{III.quad}) one can conclude that
the integrations with respect to $Q, \overline{Q}, R, R', \overline{R},
\overline{R}'$ can be done exactly as only Gaussian integrals are
involved (note, that the eigenstates of the harmonic oscillator are
Gaussian functions and derivatives of these, respectively).
Therefore, for given
$n_L, n_R, \overline{n}_L, \overline{n}_R$, one has to
perform a 6-dimensional Gaussian integral. The only difficulty lies in
the $t$-integration, which we have done numerically.
Note furthermore, that without dissipation
(i.e. $\eta=0$) the tunneling current $J$
is independent of the shift $a$, as one can show explicitly.

\section{Transition rate at resonance}
\label{sec:res}

Specifically, let us restrict ourselves to the case of zero temperature
where only the lowest level of the left well is expected to contribute
to the transition rate. Moreover, we will assume that the bias $\Delta
\approx \Omega$ and that the dissipation
leads only to a moderate broadening of the energy levels of the oscillator.
Then, one needs to consider only the levels
$n_L = \overline{n}_L = 0$ and $n_R = \overline{n}_R = 1$.

\begin{figure}
\centerline{\psfig{figure=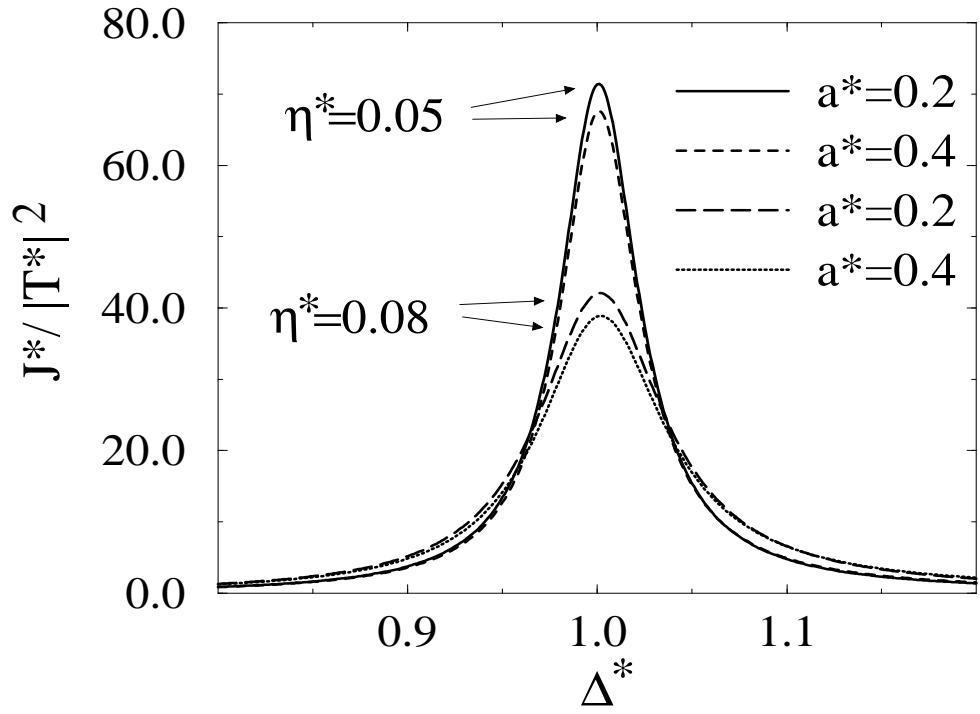,height=10cm,width=13cm}}
\vspace{1cm}
\caption{Transition rate $J^\ast/|T^\ast_{01}|^2$ as a function of the bias
$\Delta^\ast$ for $T=0$ (dimensionless, $\omega^\ast_c=60$)}
\label{trans}
\end{figure}

The calculation of the transition rate for this case is deferred to appendix
\ref{app:six}. We want to mention, however,
that the cutoff frequency $\omega_c$
of the bath spectrum is important to avoid a divergency of $\ddot{D}(t)$
for $t\rightarrow 0$. In fig. \ref{trans}, we show the results for
the transition rate calculated for different values of the
dissipation $\eta$ and the shift $a$. There, we have used dimensionless
variables which means that the frequencies and lengths were normalized
to their characteristic scales $\Omega=\sqrt{2e I_J/\hbar C}$ and
 $\kappa^{-1}=\sqrt{\hbar/M\Omega}$, respectively. The dimensionless parameters
are then given by $a^\ast = \kappa a$ for the shift and
$\eta^\ast = 1/(RC\Omega)$ for the viscosity.

In a typical experimental situation, one has $I_J\approx 1 \mbox{nA}$ for the
Josephson current and $C \approx 1 \mbox{fF}$ for the capacity. Thus, we have
$\kappa^{-1}\approx 4.2$, which means that $a^\ast \simeq \pi \kappa \approx
0.75$. The dimensionless viscosity is given by $\eta^\ast \approx 18
\mbox{k}\Omega / R$. Note, however, that the transition rate to be observed
experimentally will be very small due to the exponentially small factor
$|T_{01}|^2/(\hbar\Omega)^2$ \cite{Devoret}.

\begin{figure}
\centerline{\psfig{figure=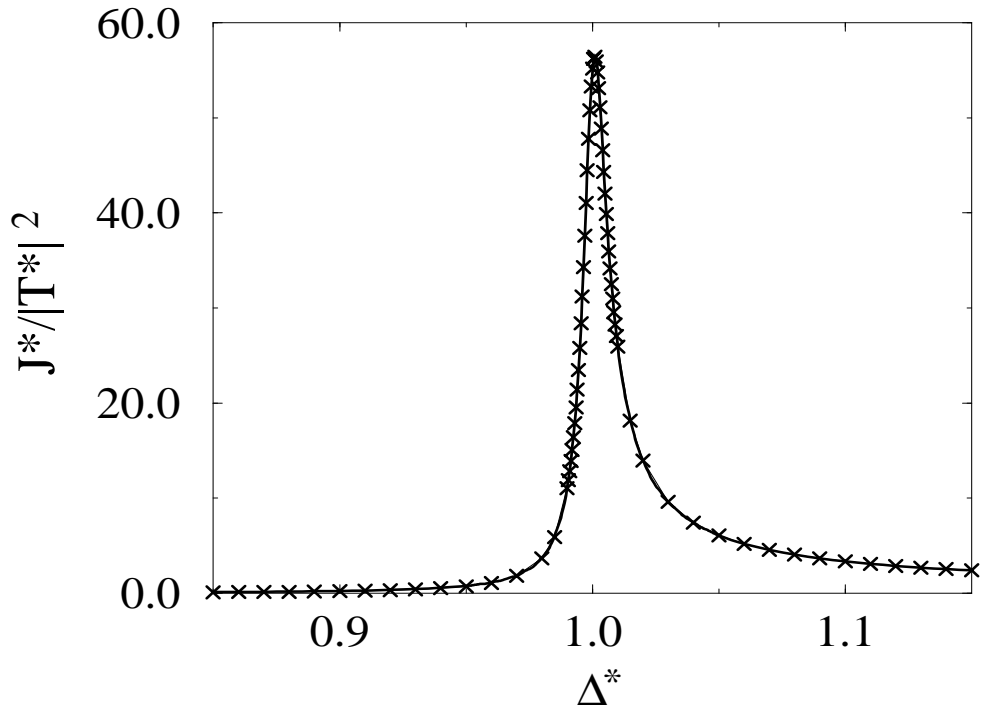,height=10cm,width=13cm}}
\vspace{1cm}
\caption{Comparison of $J^\ast(\Delta^\ast)/|T^\ast_{01}|^2$ (solid line) with
perturbation theory (crosses) (dimensionless, $\eta^\ast=0.01$, $a^\ast=4$,
$\omega^\ast_c=80$, $T=0$)}
\label{cmp}
\end{figure}

For small values of the dissipation or rather of $\alpha = 2 \eta a^2 /\pi$,
the perturbation theory \cite{Schmid94} yields the expression
\begin{equation}
\label{OS}
   J = 2 |T_{01}|^2 \,\mbox{Re}\, \left(
   e^{i\pi\alpha} \left(\frac{\gamma+i\epsilon}{\omega_c}\right)^{2\alpha}
   \frac{1}{\gamma+i\epsilon}\right),
\end{equation}
where $\gamma=\eta/(2m)$ and $\epsilon=\Delta-\Omega$. When normalized to
the same value of the peak maximum, the curves calculated from the above
formula (\ref{OS}) are nearly identical with those we obtained from our theory
(cf. fig. \ref{cmp}).
The different normalization stems from \cite{Yuri}
the way the divergencies were removed in \cite{Schmid94}.

\section{Conclusion}
\label{sec:conc}

In summary, we have studied the transition rate in a double-well system
under the influence of a dissipative environment. The rate
shows asymmetric peaks when the levels of the wells approach each other.
Thus, we have calculated the I-V characteristics of a Josephson junction
shunted by an Ohmic resistor for resonance conditions. We have modeled
the system by two parabolic potentials coupled to a bath of harmonic
oscillators. The coupling of the wells has been realized by a tunneling
Hamiltonian. Moreover, the analysis has taken into account that the
transfer of the object between the wells is a multiphonon process.
In addition, our calculation has the advantage that the bath coordinates
do not appear explicitly any more when the thermal average is taken.

\acknowledgments

The authors would like to thank A.I. Larkin and Yu.N. Ovchinnikov
for helpful discussions.

\appendix

\section{Green's functions for the harmonic oscillator in a dissipative
environment}
\label{app:green}

 From the equations of motion of the operators $\hat{R}_k (t)$ and
$\hat{p}_k (t)$ in the Heisenberg representation, one can immediately
infer the form of the retarded Green's function
$D^R (t-t') = -i \Theta(t-t') \langle[\hat{x}(t),\hat{x}(t')]\rangle$:
\begin{eqnarray}
\label{A:1}
   {(D^R)}^{-1} (\omega) &=& m [(\omega+i0)^2 - \Omega^2] - \frac{2}{\pi}\,
   \int_0 d\omega' \frac{\omega^2}{(\omega+i0)^2-{\omega'}^2}\,
   \frac{J(\omega')}{\omega'}\\
   &=& m [(\omega+i0)^2 - \Omega^2] + i\eta\omega
\end{eqnarray}
if the spectral density \cite{Leggett87} is given by
\begin{equation}
   J(\omega) = \frac{\pi}{2} \sum_j m \omega_j^3 \,\delta
   (\omega-\omega_j) = \eta\omega\,,\quad 0\leq\omega\leq\omega_c\,.
\end{equation}

In order to avoid divergencies of $\ddot{D}(t\rightarrow 0)$, it is
necessary to introduce explicitly a cutoff for $J(\omega)$ with the
characteristic frequency $\omega_c$. We used a Lorentzian cutoff
which permits to evaluate the integral in (\ref{A:1}) exactly.
For $\eta/(2m) \ll \Omega \ll \omega_c$ we obtain
\begin{eqnarray}
   m D(t) & = & i\, \mbox{Im}\, (A e^{-i\tilde{\Omega}t}) - i\, \mbox{Im}\,A\,
                e^{-t (\omega_c - \overline{\eta}/m)} \nonumber\\
                & & - \frac{1}{\pi} \,\mbox{Im}\, \left[ A \,(g(-\tilde{\Omega}
                t) + g(\tilde{\Omega}t)) \right] \nonumber\\
                & & + \frac{2}{\pi}\, \mbox{Im}\, A\; \mbox{Re}\,
                g (i\omega_ct - i\overline{\eta}t/m)\,,
\end{eqnarray}
where we have introduced
\begin{eqnarray}
   A &=& \frac{\tilde{\Omega}+i\omega_c}{2\overline{\Omega}\,
         (\tilde{\Omega} + i\omega_c - i\overline{\eta}/m)}\,,\\
   \tilde{\Omega} &=& \overline{\Omega} - i\overline{\eta}/(2m)\,,\\
   \overline{\Omega} &=& \Omega \left( 1 + \frac{\eta}{2m}
   \frac{\omega_c}{\omega_c^2+\Omega^2} \right)\,,\\
   \overline{\eta} &=& \eta \left( 1 - \frac{\Omega^2}{\Omega^2+\omega_c^2}
   \right)\,.
\end{eqnarray}
The function $g(\omega)$ is related to the exponential integral \cite{ASt}.

Finally, the correlation function $D^K(t-t') = -i \langle[\hat{x}(t),
\hat{x}(t')]_+\rangle$ is given by
\begin{equation}
	D^K_\omega = 2 i \coth{\frac{\hbar\omega}{2k_BT}}\,
        \mbox{Im}\, D^R_\omega\,,
\end{equation}
so that we have
\begin{equation}
	2 D (t) = D^R(t) + D^R(-t) + D^K(t)\,,
\end{equation}
which immediately leads to eq. (\ref{III.D}).

\section{Evaluation of the 6-dimensional Gaussian integral}
\label{app:six}

\noindent
First, let us write the transition rate (\ref{III.res}) in the form
\begin{equation}
   J = 2 \,\mbox{Re} \,\int_0 dt\, e^{i\Delta t} K(t)\,.
\end{equation}
Also, we introduce new integration variables $\bbox{x} = (v,q,u,v',q',u')$
such that
\begin{eqnarray}
   R &=& (u-v-u'+v')/4 + a\,,\nonumber\\
   R' &=& (u+v-u'-v')/4 + a\,,\nonumber\\
   \overline{R} &=& (u+v+u'+v')/4 + a\,,\nonumber\\
   \overline{R}' &=& (u-v+u'-v')/4 + a\,,\nonumber\\
   Q &=& (q-q')/2\,,\nonumber\\
   \overline{Q} &=& (q+q')/2\,.
\end{eqnarray}
In addition, the eigenstates of the harmonic oscillator are given by
\begin{eqnarray}
   \varphi_0 (R) &=& \sqrt{\kappa} \pi^{-1/4} e^{-\kappa^2 R^2/2}\,,\\
   \varphi_1 (R) &=& \sqrt{2} R \kappa\, \varphi_0 (R)\,,\quad \mbox{with}\\
   \kappa &=& \sqrt{m\Omega}\,.
\end{eqnarray}
Thus, the expression for $K(t)$, $t\ge 0$, reads
\begin{eqnarray}
   K(t) &=& \int \frac{dq dq'}{4\pi^2} du du' dv dv'\,
   |T_{01}|^2 \,\frac{\kappa^2}{16 \pi} \nonumber\\
   & & \times\, \left.\frac{\partial^2}{\partial\overline{a}
   \partial\overline{a}'}\; \exp{(-\bbox{x M x} + \bbox{b x} + c)}
   \right|_{\overline{a}=0=\overline{a}'}\,.
\label{B:1}
\end{eqnarray}

The only nonzero elements of the symmetric $6\times 6$
matrix $\bbox{M}$ are given by
\begin{eqnarray}
   M_{11} &=& \frac{\kappa^2}{4} - \frac{i m^2}{2} (\ddot{D}(0) -
      \ddot{D}(t))\,,\nonumber\\
   M_{22} &=& \frac{i}{2} (D(0)+D(t))\,,\nonumber\\
   M_{33} &=& \frac{\kappa^2}{4} = M_{66}\,,\nonumber\\
   M_{12} &=& M_{21} = - \frac{im}{2} \dot{D}(t) = -M_{45} = -M_{54}
     \,,\nonumber\\
   M_{23} &=& M_{32} = - \frac{i}{4} = M_{56} = M_{65}\,,\nonumber\\
   M_{44} &=& \frac{\kappa^2}{4} - \frac{i m^2}{2} (\ddot{D}(0) +
      \ddot{D}(t))\,,\nonumber\\
   M_{55} &=& \frac{i}{2} (D(0)-D(t))\,.
\end{eqnarray}
Thus, it can be decomposed into two $3\times 3$ blocks.
Furthermore, the vector $\bbox{b}$ reads
\begin{eqnarray}
   \bbox{b} &=& \frac{i\kappa\overline{a}}{4} (1,0,1,1,0,1) -
   \frac{i\kappa\overline{a}'}{4} (1,0,1,-1,0,-1)\nonumber\\
    & & + \left(\kappa^2 a - 2i m^2\Omega^2 a [D(t)-D(0)],\; ia +
   2i m\Omega^2 a \int_0^t dt' D(t'),\; 0,0,0,0\right)\,,
\end{eqnarray}
and
\begin{equation}
   c = -2\kappa^2 a^2 - i\kappa a (\overline{a}-\overline{a}')
      - i 2 m^2 \Omega^4 a^2 \int_0^t dt_1 dt_2 \,D(t_1-t_2)
      - i 2 m \Omega^2 a^2 t\,.
\end{equation}

The Gaussian integral in eq. (\ref{B:1}) can be evaluated exactly
and eventually, we obtain
\begin{eqnarray}
   K(t) &=& W(t) \,e^{S(t)}\,,\\
   W(t) &=& \frac{\kappa^2}{8} |T_{01}|^2 (\mbox{det} \bbox{M})^{-1/2}\,
   \nonumber\\
   & & \times\, \left\{ \frac{\partial\bbox{b}}{\partial\overline{a}}
   (\bbox{M}^{-1}) \frac{\partial\bbox{b}}{\partial\overline{a}'} -
   \left[ \bbox{b} (\bbox{M}^{-1}) \frac{\partial\bbox{b}}{\partial
   \overline{a}} + \frac{\partial c}{\partial\overline{a}}
   \right]^2 \right\}_{\overline{a}=0=\overline{a}'}\,,\\
   S(t) &=& \left\{\frac{1}{2} \bbox{b} (\bbox{M}^{-1}) \bbox{b} + c
   \right\}_{\overline{a}=0=\overline{a}'}\,.
\end{eqnarray}
In the case of zero dissipation, this expression is indeed independent
of the shift $a$, as it can be shown exactly.

\clearpage
\clearpage
\end{document}